\documentclass[sigconf]{acmart}

\usepackage{amsmath}
\usepackage{multirow}
\usepackage{balance}

\AtBeginDocument{
	\addtolength{\abovedisplayskip}{-1ex}
	\addtolength{\abovedisplayshortskip}{-1ex}
	\addtolength{\belowdisplayskip}{-1ex}
	\addtolength{\belowdisplayshortskip}{-1ex}
}

\AtBeginDocument{%
  }

\copyrightyear{2022}\acmYear{2022}\setcopyright{acmcopyright}\acmConference[RecSys '22]{Sixteenth ACM Conferenceon Recommender Systems}{September 18--23, 2022}{Seattle, WA, USA}\acmBooktitle{Sixteenth ACM Conference on Recommender Systems (RecSys '22), September 18--23, 2022, Seattle, WA, USA}\acmPrice{15.00}\acmDOI{10.1145/3523227.3551477}\acmISBN{978-1-4503-9278-5/22/09}

\begin{document}
\title[M2TRec: Metadata-aware Multi-task Transformer for Session-based Recommendations]{M2TRec: Metadata-aware Multi-task Transformer for Large-scale and Cold-start free Session-based Recommendations}

\author{Walid Shalaby}
\email{walid_shalaby@homedepot.com}
\affiliation{%
  \institution{The Home Depot}
	  \city{Atlanta}
      \state{Georgia}
 \country{USA}
}

\author{Sejoon Oh}
\email{soh337@gatech.edu}
\affiliation{%
	\institution{Georgia Institute of Technology}
	  \city{Atlanta}
      \state{Georgia}
	\country{USA}
}

\author{Amir Afsharinejad}
\email{amir_afsharinejad@homedepot.com}
\affiliation{%
  \institution{The Home Depot}
	  \city{Atlanta}
      \state{Georgia}
 \country{USA}
}

\author{Srijan Kumar}
\email{srijan@gatech.edu}
\affiliation{%
	\institution{Georgia Institute of Technology}
	  \city{Atlanta}
      \state{Georgia}
	\country{USA}
}

\author{Xiquan Cui}
\email{xiquan_cui@homedepot.com}
\affiliation{%
	\institution{The Home Depot}
	  \city{Atlanta}
      \state{Georgia}
	\country{USA}
}

\renewcommand{\shortauthors}{Shalaby et al.}


\begin{abstract}
    Session-based recommender systems (SBRSs) have shown superior performance over conventional methods. However, they show limited scalability on large-scale industrial datasets since most models learn one embedding per item. This leads to a large memory requirement (of storing one vector per item) and poor performance on sparse sessions with cold-start or unpopular items. Using one public and one large industrial dataset, we experimentally show that state-of-the-art SBRSs have low performance on sparse sessions with sparse items. We propose M2TRec, a Metadata-aware Multi-task Transformer model for session-based recommendations. Our proposed method learns a transformation function from item metadata to embeddings, and is thus, item-ID free (i.e., does not need to learn one embedding per item). It integrates item metadata to learn shared representations of diverse item attributes. During inference, new or unpopular items will be assigned identical representations for the attributes they share with items previously observed during training, and thus will have similar representations with those items, enabling recommendations of even cold-start and sparse items. Additionally, M2TRec is trained in a multi-task setting to predict the next item in the session along with its primary category and subcategories. Our multi-task strategy makes the model converge faster and significantly improves the overall performance. Experimental results show significant performance gains using our proposed approach on sparse items on the two datasets.
\end{abstract}

\ccsdesc[500]{Information systems~Recommender systems}


\maketitle

\section{Introduction}

\par Session-based recommender systems (SBRSs) accurately model sequential and evolving preferences of users from their session data (e.g., clicks and add-to-cart events). Session data can be associated with item metadata, allowing SBRSs to capture item dependencies at the attribute level within the session.
However, most of the existing SBRSs take the IDs of users and items as the main input source to learn session contexts and produce next item recommendations
\cite{liu2018stamp,wu2019session,li2017neural}. 
Recent hybrid models demonstrated improved performance when combining item embeddings and their attributes to be used as additional side information \cite{wang2017perceiving,song2021cbml,twardowski2016modelling,tuan20173d,hidasi2016parallel,
	twardowski2016modelling,de2021transformers4rec}. 

\par However, there are two major issues that these recommendation models face.  The first issue is that they cannot scale with the gigantic sizes of industrial datasets. For instance, The Home Depot (THD) industrial dataset used in this paper has approximately 40 million sessions and 0.6 million items. Since the dataset is created by sampling several months of online sessions, the actual full dataset (e.g., a year-long one) will be even much bigger. Most SBRSs~\cite{liu2018stamp, wu2019session,li2017neural, COTREC, yu2020tagnn, pan2020intent, hidasi2018recurrent, de2021transformers4rec} that utilize a large item-ID embedding matrix can suffer from slow training or memory shortage problems. 

The second issue arises due to cold-start items and sparse sessions, i.e., sessions that contain new or unpopular items. SBRSs will have limited or no ability to generate good representations for such items since they have no-to-few interactions.
Moreover, many existing models are incapable of scoring and recommending new items unseen during training~\cite{
	saveski2014item, vlachos2018addressing}.
Even combining metadata information with item-IDs, i.e., item embeddings, to learn compound item representation results in only a slight performance improvement compared to using item-ID only \cite{de2021transformers4rec,wang2017perceiving,vasile2016meta,song2021cbml,meng2020incorporating,hidasi2016parallel,tuan20173d}. This can be attributed to the model overfitting item-ID as the main feature.

\begin{figure*}[th]
	\centering
	\makebox[\linewidth]{\includegraphics[keepaspectratio=true,scale=0.1]{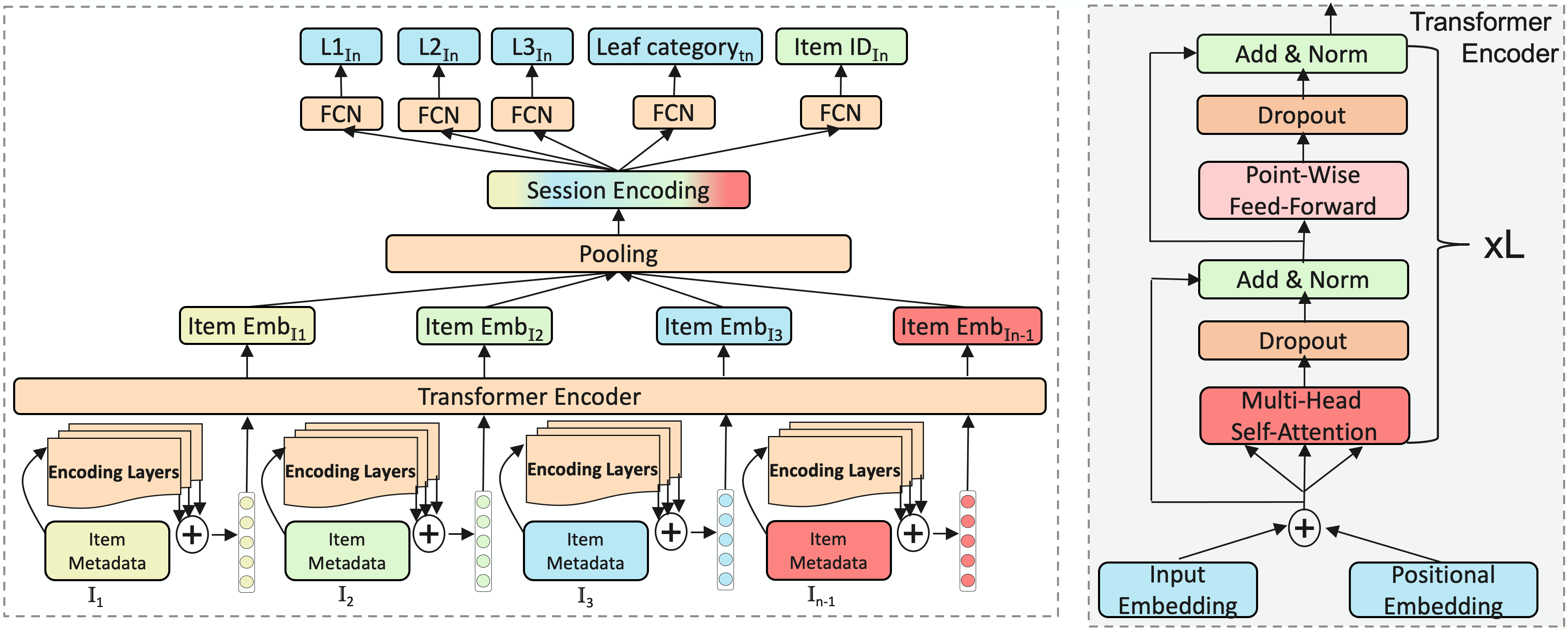}}
	\caption{M2TRec Architecture. (left) M2TRec first encodes metadata of each item in a session into embeddings and feeds the concatenation of those metadata embeddings to a Transformer encoder. A current session encoding is obtained via average-pooling of the item encodings from Transformer, and the session encoding is used for next item and next category predictions. (right) Detailed architecture illustration of a Transformer encoder used in M2TRec.}
	\label{fig:spp-arch}
\end{figure*}

\par To tackle the above issues, we propose M2TRec, a Metadata-aware Multi-task Transformer model (Figure \ref{fig:spp-arch}). M2TRec is completely item-ID free (i.e., no item-ID embeddings) and uses only item attributes such as title, category, brand, color, and other metadata to learn item representations.
Since M2TRec does \textit{not} require creating and learning a large item-ID embedding matrix, it can be easily applied to large industrial datasets.
In addition, new items will still have accurate representations using their metadata attributes, and their similarity with previously-observed items can be captured even if they have no or few interactions with users, making it suitable even in highly dynamic deployment settings where the item catalog changes frequently. 
Finally, recent research has shown that multi-task learning (MTL) results in increased performance and generalizability of the model~\cite{hadash2018rank, gao2019neural}. Thus, given the focus on leveraging item metadata, we design M2TRec as a multi-task SBRS trained not only to predict the next item but also its category and subcategories to enhance the prediction accuracy of an individual task.

Via thorough experiments, we demonstrate the superior performance of the proposed model on real-world datasets. Our multi-task training allows faster convergence, higher accuracy with fewer iterations, and robust performance with fewer training data (some experimental results are omitted due to space limits). 
Our scalable architecture serves both item and category recommendations in one model with higher prediction performance than baselines in both tasks.

\section{Related Work}
Neural networks have served as the main architecture of existing SBRSs. Earlier works~\cite{hidasi2015session,hidasi2016parallel,hidasi2018recurrent} utilize a recurrent neural network (RNN) to model sequential dependencies of items within a session. Recently, attention-based approaches~\cite{li2017neural,liu2018stamp, wang2019collaborative, pan2020intent, ren2019repeatnet} and graph neural network (GNN)-based methods~\cite{yu2020tagnn, COTREC, wu2019session, xu2019graph, wang2020global} have been proposed to enhance the longer and deeper dependencies for SBRSs.
Furthermore, Transformer~\cite{vaswani2017attention}-based SBRSs~\cite{de2021transformers4rec, de2021end, chen2019bert4sessrec, moreira2021transformers} show the state-of-the-art prediction performance due to its powerful and efficient self-attention mechanism.
Multi-task learning (MTL)~\cite{qiu2021incorporating, huang2021graph, meng2020incorporating} has been also adopted for SBRSs to  enhance the next item prediction via generalization. However, the above models have shortcomings that they (1) are susceptible to cold-start items or sessions, (2) cannot predict the categories of the next item in a session, or (3) are not scalable to the real-world billion-scale recommendation setting since they have to store item-IDs and their embeddings, which are this paper is solving.

Methodologies are developed to incorporate item metadata into SBRSs for modeling user/item dependencies \cite{song2016multi,jannach2017session,you2019hierarchical, de2021transformers4rec,wang2017perceiving,vasile2016meta,song2021cbml,meng2020incorporating,qiu2020exploiting,hidasi2016parallel,tuan20173d}. However, the vast majority of such methodologies have at least one of the following two shortcomings. 
The first shortcoming is that cold-start items and/or users are removed during the pre-processing of datasets used for evaluations of proposed models~\cite{song2016multi,jannach2017session,you2019hierarchical,qiu2020exploiting, de2021transformers4rec}. 
Item metadata alone cannot give the model the ability to recommend such cold-start items \cite{moreira19on}. The model needs to have a separate mechanism for using items' content information and representation to recommend cold-start items.  \citet{tagliabue20the} propose a pipeline to learn accurate cold-start item representations with small changes to an existing model infrastructure. However, a separate neural model needs to be trained to obtain the cold-start embeddings,  which limits the scalability of the solution. \citet{raziperchikolaei2021shared} and \citet{zheng2021cold} suggest hybrid and metadata-aware recommendation models to predict implicit feedback for cold-start items of users, respectively. 
However, those models are not designed for the session-based recommendation setting. 
The second shortcoming of such methodologies is that they do not make use of item titles or descriptions as attribute features for capturing product similarities~\cite{wang2017perceiving,vasile2016meta,song2021cbml,meng2020incorporating,jannach2017session,qiu2020exploiting}. In such cases, item-IDs are used as inputs to represent different products. Such representation is unable to incorporate any relevant information about cold-start items as opposed to using title encodings.

\begin{table*}[t] \label{tbl:1}
	\small
	\centering
	\caption{Statistics of datasets used in the experiments.}
	\begin{tabular}{|l|c|c|c}
		\hline
		Dataset  & Diginetica & THD \\
		\hline
		\# of training and test sessions & 191K, 16K & 39M, 1.5M \\
		\hline
		\# of items & 119K & 575K \\
		\hline
		Metadata / Attributes & \shortstack[c]{Product title, Category} & \shortstack[c]{Product title, Categories (L1, L2, L3, Leaf), \\ Manufacturer, Brand, Department Name, \\ Class Name, Color} \\
		\hline
		Prediction tasks & \shortstack[c]{Item-ID,  Category} & \shortstack[c]{Item-ID, Categories (L1, L2, L3, Leaf)} \\
		\hline
	\end{tabular}
	\label{tab:dataset}
\end{table*}

\section{Methodology}
\noindent \textbf{Next Item Prediction:}
We denote a user session $\mathcal{S}=[I_{1}, I_{2}, I_{3}, ..., I_{n}]$ as a sequence of items a user interacted within that session. Each item 
\begin{math}I_k=\{A_{k,1}, A_{k,2}, A_{k,3}, ..., A_{k,m}\}\end{math} is described by a set of \begin{math}m\end{math} attributes which could be context-specific or item-specific. In this work, we consider item-specific attributes only (e.g., title, description, category). Each attribute $A$ could be either textual, categorical, or numerical. 
In the setting of session-based recommendations, we are given a session $\mathcal{S}$, and our objective is to maximize the prediction probability of the next item the user is most likely to interact with given all previous items in $\mathcal{S}$. Formally, the probability of the target item $I_n$  can be formulated as:
\begin{equation} \label{eq:1}
	p(I_n|\mathcal{S}_{[I_{<n}]}; \theta)
\end{equation}	

where $\theta$ denotes the model parameters and $\mathcal{S}_{[I_{<n}]}$ denotes the sequence of items prior to the target item $I_n$. 
As in previous works~\cite{hidasi2015session, liu2018stamp, li2017neural, wu2019session}, we generate dense next item sub-sequences from each session $\mathcal{S}$ for training and testing. Therefore, a session $\mathcal{S}$ with $n$ items will be broken down into $n-1$ sub-sequences such as $\{([I_{1}], I_{2}), ..., ([I_{1}, I_{2}, ..., I_{n-1}], I_{n})\}$, where ([X], Y) means X as the input sequence of items and Y as the target next item.

\vspace{1mm}
\noindent  \textbf{Item Metadata Encoding:} \label{sec:item-meta-encoding}
Item metadata can be numerical, categorical, or unstructured such as title, description, and image. We propose a unified method for representing all item attributes. The objective is to map every attribute $A$ into a real-valued vector $v_{A} \in \mathbb{R}^{d_{{A}}}$. 
Numerical attributes $r$ are represented as a single-valued vector $v_r \in \mathbb{R}$. 
Categorical attributes $C \in \{c_1, c_2, ..., c_s\}$ are encoded into vectors $v_C$ using an embedding layer dedicated to each attribute, i.e.,
\begin{equation} \label{eq:2}
	v_{C} = c_i \theta^{(C)}  \in \mathbb{R}^{d_{C}}
\end{equation}

where $c_i$ is the one-hot encoded value of $C$, $\theta^{(C)} \in \mathbb{R}^{s \times d_{C}}$ are the weights of the category embedding matrix, $s$ is the number of possible values of $C$, and $d_C$ is the dimensionality of $C$'s vector. 

Textual attributes $T$ are first tokenized using a subword tokenizer \cite{sennrich2015neural} to obtain individual tokens $[\textbf{w}_{1}, \textbf{w}_{2}, ..., \textbf{w}_{t}]$ and then encoded into vectors $v_{T}$. 
A simple and efficient encoding strategy is to create a dedicated embedding layer for $T$ to map each token $\textbf{w}$ into a vector and then aggregate the token vectors using mean or max pooling, i.e., 
\begin{equation} \label{eq:3}
	v_{T} = \text{Pool}_{i=1}^t(w_i \theta^{(T)})  \in \mathbb{R}^{d_{T}}
\end{equation}
where $w_i$ is the one-hot encoded value of token $\textbf{w}_i$, $\theta^{(T)} \in \mathbb{R}^{k \times d_{T}}$ are the weights of the token embedding matrix, $k$ is vocabulary size of $T$, and $d_T$ is the dimensionality of $T$'s vector. 
Further enhancements to textual attributes encoding can be achieved by sharing the encoding parameters across all textual attributes that have similar vocabularies such as item title, description, category, color, etc. 
Although the weight-sharing scheme is expected to reduce the training time, it may increase the overall model size. This is because the vector size would be the same for all the attributes that share the same encoder regardless of their vocabulary size. This will lead to high memory and storage requirements when deploying the model in production. Alternatively, in this work, we use a separate embedding layer for each textual attribute as in Eq. (\ref{eq:3}) and choose its vector size proportional to the attribute's vocabulary size.  

\vspace{1mm}
\noindent \textbf{Session Encoding:}
After encoding all metadata features for an item $I_k$ at position $k$ in the input session $S$, we concatenate all the feature vectors to create a compound vector representation $v_{I{_k}}$ for $I_k$, i.e., 
\begin{equation} \label{eq:5}
	v_{I{_k}} = concat(v_{A_1}, v_{A_2}, ..., v_{A_m})  \in \mathbb{R}^{d_I}
\end{equation}
where $d_I$ is the summation of the lengths of all feature vectors.
Note that item-IDs are not used to create the compound representation $v_{I{_k}}$ in Equation~\eqref{eq:5}. 
We then use the compound representations of  items in $S$ as input to the session encoder in a pre-fusion fashion to learn a session encoding $v_S$. First, Transformer encoder~\cite{vaswani2017attention} generates contextual encodings for each session item $v_{I_k}$, followed by an average-pooling layer to generate the session encoding $v_S$, i.e., 
\begin{equation} \label{eq:6}
	v_{S} = \text{Pool}_{k=1}^{n-1}(\text{Trans-Enc}(v_{I_{k}}; \theta^{(enc_{S})}))  \in \mathbb{R}^{d_{S}}
\end{equation}
where $\theta^{(enc_S)}$ is the model parameters of the Transformer encoder trained with sessions $S$.

\vspace{1mm}
\noindent \textbf{Multi-task Learning:}
M2TRec incorporates multi-task learning (MTL) to boost the performance of next item prediction. MTL has proven to be an effective mechanism to reduce the risk of overfitting, learn more generalized shared representations for all the tasks, and improve the overall performance on each task by sharing the knowledge acquired from other related tasks \cite{ruder2017overview}.  
The target space of next item prediction (i.e., all item-IDs) is much larger than the space of other item attributes such as all categories or brands. Therefore, the task of next item category or brand prediction should be easier to learn than next item prediction. Moreover, learning such auxiliary tasks would benefit the task of next item prediction since it biases the metadata encoding and the Transformer encoder layers to learn representations that are close not only to next item, but also to other similar items belonging to the same category or brand, thus, narrowing down the space of possible next item candidates to a  much smaller set of items. 
To this end, we train M2TRec to predict  next item attributes (e.g., item categories) as auxiliary tasks to the task of next item-ID prediction. For each task, including next item-ID prediction, we create a prediction head composed of Fully Connected Layer (FCN) followed by Softmax function to generate the probability distribution over all candidates for the corresponding task, i.e., 
\begin{equation} \label{eq:7}
	\hat{y}_k = softmax(\text{FCN}_k(v_S; \theta^{(k)}))  \in \mathbb{R}^{d_k}
\end{equation}
where $\theta^{(k)}$ is the parameters of the FCN for the $k^{\text{th}}$ task, and $d_{k}$ is the total number of possible outputs of the $k^{\text{th}}$ task (e.g., total number of categories for category predictions).

The loss of the $k^{\text{th}}$ task prediction head and overall prediction of M2TRec are calculated using cross-entropy loss as follows, respectively: 
\begin{equation} \label{eq:8}
\mathcal{L}_{k} = -\sum_{i=1}^{\mathrm{d_k}} y_{k_i} \cdot \mathrm{log}\; {\hat{y}}_{k_i}, \qquad 	\mathcal{L} = \sum_{k=1}^{\mathrm{N}} \mathcal{L}_k
\end{equation}
where $y_{k}$ is a one-hot encoding including the ground-truth information for the $k^{\text{th}}$ task.

\begin{table*}[t]
	\small
	\centering
	\caption{The performance of M2TRec on next item prediction task on all sessions (All) and sessions with tail items (Sparse) compared to other baseline methods. 
		Tail items indicate items with less than 10 occurrences in a dataset.
		(\textbf{Bold} indicates the best model, while the second-best model is \underline{underlined}).}
	\label{tbl:performance-on-all-tail-sessions}
	\begin{tabular}{|l|c|c|c|c|c|c|c|c|c|c|c|c|}
	\hline
	Dataset  & \multicolumn{6}{c|}{Diginetica} & \multicolumn{6}{c|}{THD} \\
	\hline
	\multirow{2}{0cm}{Method}  & \multicolumn{2}{c|}{HIT@20} & \multicolumn{2}{c|}{Recall@20} & \multicolumn{2}{c|}{MRR@20} & \multicolumn{2}{c|}{HIT@20} & \multicolumn{2}{c|}{Recall@20} & \multicolumn{2}{c|}{MRR@20} \\
	\cline{2-13}
	& All & Sparse & All & Sparse & All & Sparse & All & Sparse & All & Sparse & All & Sparse \\
	\hline
	STAMP & 25.45 & 16.05 & 41.61 & 28.89 & 7.17 & 4.44 & 31.91 & 14.65 & \underline{38.49} & 18.90 & 14.88 & 6.46 \\
	GRU4Rec & 29.09 & \underline{22.33} & 45.45 & 36.14 & \underline{8.51} & \underline{6.45} & \underline{32.27} & \underline{16.81} & 37.64 & \underline{20.28} & \underline{14.97} & \textbf{7.93} \\
	NARM & \underline{30.49} & 21.78 & \underline{47.60} & \underline{36.90} & 8.13 & 5.55 & 28.98 & 11.31 & 35.45 & 15.59 & 11.63 & 3.97 \\
	\hline
	M2TRec & \textbf{35.47} & \textbf{25.51} & \textbf{53.70} & \textbf{41.01} & \textbf{9.75} & \textbf{6.63} & \textbf{34.78} & \textbf{19.05} & \textbf{41.51} & \textbf{24.86} & \textbf{15.65} & \underline{7.38} \\
	Improvement & 16.33\% & 14.2\% & 12.82\% & 11.1\% & 14.57\% & 2.79\% & 7.78\% & 13.3\% & 7.85\% & 22.6\% & 4.54\% & -6.94\% \\
	\hline
\end{tabular}
\end{table*}

\section{Experiments}

\noindent \textbf{Datasets:} We conducted our experiments on two real-world datasets. 
We exclude all sessions with only one item.
Table \ref{tab:dataset} shows dataset statistics along with the item metadata we used and the prediction tasks for each dataset.

\begin{itemize}
	\item \textit{Diginetica} 
	\footnote{https://competitions.codalab.org/competitions/11161}
	is an E-commerce dataset that was a part of CIKM Cup 2016 challenge. We use the transactional and product data and use pre-processing similar to \cite{wu2019session}.

	\item \textit{THD} is an E-commerce dataset obtained from The Home Depot, the largest home improvement retailer in the USA.
	The dataset is composed of Add-to-Cart (ATC) events within millions of online sessions.  Similar to SIGIR 2021 data challenge dataset\cite{tagliabue2021sigir}, training data is created by sampling several months of online purchase sessions. Test data is sampled from a disjoint and adjacent time period. The dataset has rich product metadata including 7 attributes: product title, categories (L1, L2, L3, Leaf), brand, manufacturer, color, department, and class name. 
\end{itemize}

\vspace{1mm}
\noindent \textbf{Baselines:}
We select the following 3 state-of-the-art SBRSs to compare them with M2TRec: (1) 
\textit{\textbf{GRU4Rec}}~\cite{hidasi2015session}: A popular and first-generation SBRS that utilizes a Gated Recurrent Unit (GRU)~\cite{cho2014properties} to model long-term dependencies within a session,
(2) \textit{\textbf{NARM}}~\cite{li2017neural}: An attention-based SBRS that employs a hybrid encoder to reflect a user's global and local interests with an attention mechanism, and
(3) \textit{\textbf{STAMP}}~\cite{liu2018stamp}: An attention/memory-based SBRS that incorporates a user's short-term and long-term interests via a short-term attention and long-term memory modules, respectively.

\begin{figure*}[th] 
	\centering
	\includegraphics[keepaspectratio=true,scale=0.3]{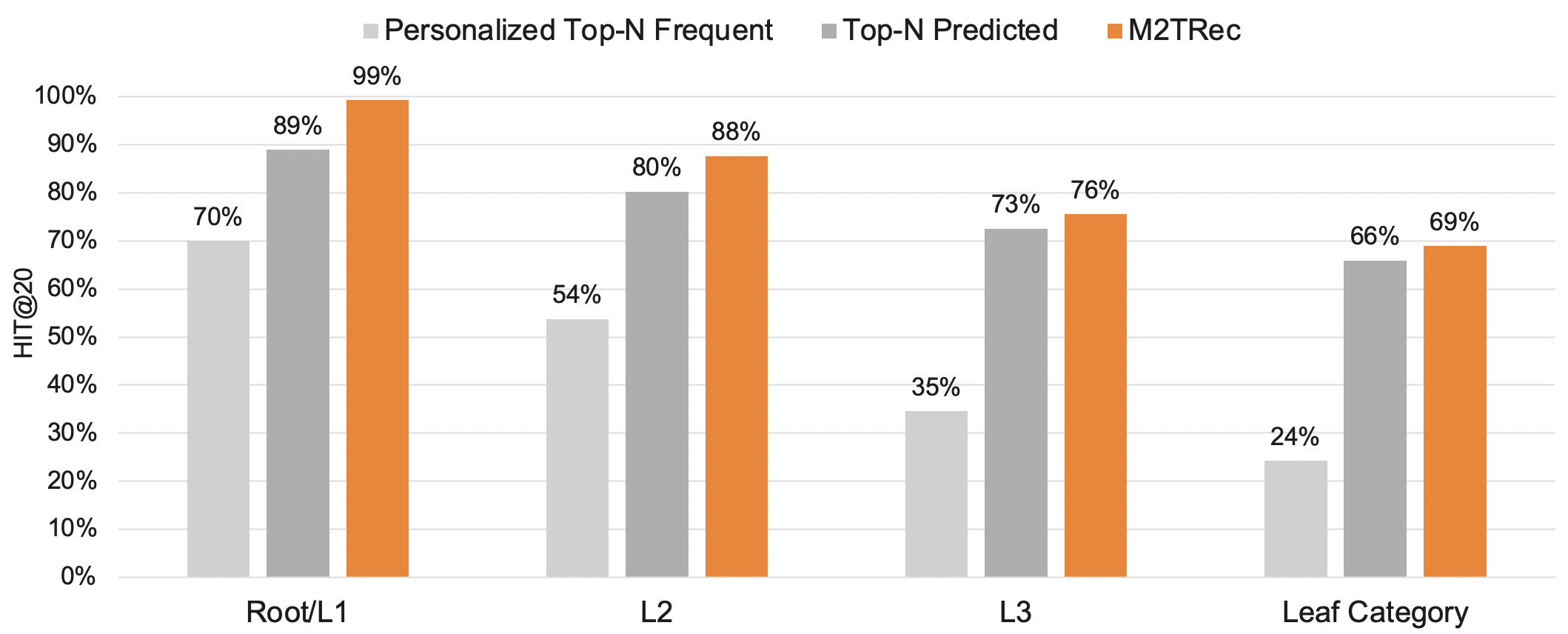}
	\caption{M2TRec Performance on Category Recommendation}
	\label{fig:category-rec}
\end{figure*}

\begin{table*}[t]
	\small
	\centering
	\caption{The performance of M2TRec on all sessions (All) and sessions with tail items (Sparse) compared to variants of M2TRec.}
	\label{tbl:ablation-performance-on-all-sessions}
	\begin{tabular}{|l|c|c|c|c|c|c|c|c|c|c|c|c|l|}
	\hline
	Dataset  & \multicolumn{6}{c|}{Diginetica} & \multicolumn{7}{c|}{THD} \\
	\hline
	\multirow{2}{0cm}{Method}  & \multicolumn{2}{c|}{HIT@20} & \multicolumn{2}{c|}{Recall@20} & \multicolumn{2}{c|}{MRR@20} & \multicolumn{2}{c|}{HIT@20} & \multicolumn{2}{c|}{Recall@20} & \multicolumn{2}{c|}{MRR@20} & \multirow{2}{0cm}{Model Size}\\
	\cline{2-13}
	& All & Sparse & All & Sparse & All & Sparse & All & Sparse & All & Sparse & All & Sparse & \\
	\hline
	TRec$_{id}$ & 30.24 & 16.93 & 46.96 & 29.45 & 8.55 & 4.35 & 32.90 & 11.58 & 39.40 & 15.22 & 15.31 & 4.90 & 110.3M \\
	\hline
	MuTRec$_{id}$ & 32.69 & 21.10 & 50.13 & 35.18 & \underline{9.50} & 5.71 & \underline{33.73} & 13.93 & \underline{40.35} & 17.92 & \underline{15.43} & 5.90 & 111.1M \\
	\hline
	TRec$_{title}$ & 32.17 & 20.41 & 49.40 & 34.64 & 8.70 & 5.22 & 32.29 & 16.48 & 38.64 & 21.19 & 14.25 & 6.32 & 56.1M \\
	\hline
	MeTRec & \underline{34.78} & \underline{24.99}& \underline{52.50} & \underline{40.43} & 9.33 & \underline{6.22} & 33.54 & \underline{17.10} & 40.04 & \underline{21.97} & 15.17 & \underline{6.42} & 86.5M \\
	\hline
	M2TRec & \textbf{35.47} & \textbf{25.51} & \textbf{53.70} & \textbf{41.01} & \textbf{9.75} & \textbf{6.63} & \textbf{34.78} & \textbf{19.05} & \textbf{41.51} & \textbf{24.86} & \textbf{15.65} & \textbf{7.38} & 89.5M \\
	\hline
\end{tabular}
\end{table*}

\vspace{1mm}
\noindent  \textbf{Evaluation Metrics:} 
We use HIT@K, Recall@K, and MRR@K~\cite{voorhees1999trec} to evaluate the performance of M2TRec.
All the metrics range from 0 to 1, and higher values are better. We choose $K=20$ since it is a standard value~\cite{ludewig2019performance}.

\vspace{1mm}
\noindent \textbf{Implementations:} We used open-source implementations for all baseline methods\footnote{https://github.com/rn5l/session-rec}. 
With M2Trec, we encode all the attributes as textual. We use a dedicated embedding layer for each attribute followed by average pooling of individual tokens' vectors. The embedding dimension is set proportionally to the total number of distinct tokens of the corresponding attribute vocabulary.  
For the Transformer encoder, we used 2 encoder layers with 8 attention heads in each layer and point-wise feed-forward networks consisting of two fully-connected layers [2048, 128] with a ReLU activation \cite{agarap2018deep}  in between.
We fine-tuned all the hyperparameters of M2TRec on a validation dataset sampled randomly from THD data. 

\vspace{1mm}
\noindent \textbf{Next item prediction task on all sessions:}
The performance of all the models on all sessions of the two datasets is shown in Table \ref{tbl:performance-on-all-tail-sessions}.
As we can notice, M2TRec outperforms all the baselines across all the evaluation metrics. On Diginetica, the relative performance improvements of HIT@20, Recall@20, and MRR@20 are in the range of 13\% $\sim$ 16\%. On THD dataset, the relative performance improvements are in the range of 5\% $\sim$ 8\%. As in \cite{song2021cbml}, we compute the relative performance improvement of a metric as the difference in the performance of M2TRec and the second runner over the performance of the second runner on that metric reported in percentage. These improvements indicate the effectiveness of utilizing item metadata and the multi-task learning regime which are unique to M2TRec, compared to other baselines which use only item-ID as the main and only input for session-based recommendations.

\vspace{1mm}
\noindent \textbf{Next item prediction task on sparse sessions:} Table \ref{tbl:performance-on-all-tail-sessions} highlights the performance of all the models on sparse sessions containing cold-start or tail items in the two datasets. Tail items indicate items with less than 10 occurrences in a dataset. These sessions represent 34\% and 12\% of the total sessions in Diginetica and THD datasets respectively. As we can notice, M2TRec relative improvements on sparse sessions are much higher than all the other models, especially on HIT@20 and Recall@20 for both datasets (e.g., 11\% $\sim$ 23\% boost on both datasets). These results demonstrate the effectiveness of our proposed item-ID free approach on sparse sessions and its robustness in mapping tail and cold-start items within these sessions into meaningful representations based on their metadata.

\vspace{1mm}
\noindent \textbf{Predicting next item's category:}
One of the main objectives of this research is to develop a scalable architecture that serves both item and category recommendations in one model using an efficient MTL regime. We found significant performance gains when jointly training our model to predict next item and its categories at different levels of the catalog taxonomy (see ablation study below). 
We demonstrate the efficacy of training our SBRS to predict next category over deriving it from session items by comparing the category prediction performance against two heuristics:
(1) \textit{\textbf{Personalized top-N Frequent:}} This simple heuristic uses past session items' categories and recommends the most frequent ones, and (2) \textit{\textbf{Top-N Predicted:}} This simple strategy works by first predicting top-N next items from a metadata-aware single task model called MeTRec (see ablation study below), and then uses their categories as recommendations such that the category of the top-ranked next item will be ranked first and so on.

Figure \ref{fig:category-rec} shows the performance of category recommendation using M2TRec against the two heuristics. Performance is measured in terms of HIT@20 and reported for L1, L2, L3, and leaf categories of THD dataset. As we can notice, the performance of M2TRec is significantly better than the two other strategies across all tasks. For example, on leaf category prediction, the performance gains are in the range of 3\%$\sim$45\%. These results demonstrate that M2TRec can effectively perform next category prediction tasks.

\vspace{1mm}
\noindent \textbf{Ablation Study of M2TRec:}
To investigate the contribution of each component of  M2TRec, we developed the following variants: (1) \textbf{\textit{TRec}}$_{title}$: A variant of M2TRec which uses item title as the only input feature without any additional meta-data. It leverages the same architecture in Figure \ref{fig:spp-arch}, but has one prediction head only to predict next item-ID, (2) \textbf{\textit{MeTRec:}} Metadata-aware variant which utilizes all metadata as input features. This variant also has one prediction head to predict next item-ID, 
(3) \textbf{\textit{TRec}}$_{id}$: A variant of M2TRec which uses item-IDs as the only input feature without any additional meta-data attributes. It leverages the same architecture in Figure \ref{fig:spp-arch} but has only one prediction head to predict next item-ID, and
(4) \textbf{\textit{MuTRec}}$_{id}$: Multi-task variant of TRec$_{id}$. The model is trained on the same tasks as M2TRec, but uses only item-IDs as the input features (i.e., it does not use other metadata features).

The performance of M2TRec and its variants on all and sparse sessions is shown in Table \ref{tbl:ablation-performance-on-all-sessions}, where model size indicates the number of model parameters.  M2TRec outperforms all other variants significantly, especially on sparse sessions where the performance gains on Diginetica dataset are in the range of 1\%$\sim$9\% in terms of HIT@20 and 1\%$\sim$12\% in terms of Recall@20. On THD dataset, performance gains are in the range of 2\%$\sim$7\% and 3\%$\sim$10\% in terms of HIT@20 and Recall@20 respectively. 
As we include all metadata in MeTRec, the performance on sparse sessions outpaces all other variants. Moreover, the performance on all sessions improves significantly and outpaces TRec$_{id}$ and its multi-task version (MuTRec$_{id}$) on Diginetica, while it is on par with MuTRec$_{id}$ on THD dataset. This demonstrates the usefulness of metadata-awareness and its sufficiency in providing competitive performance to classical item-ID based SBRSs. As we can notice, M2Trec and MeTRec are about 19\%-22\% less in size than the item-ID based variants. Besides, all the metadata-aware variants are more scalable to the increase in item catalog size compared to the item-ID based variants.

\section{Conclusion}
This work provides a scalable and practical solution for leveraging metadata to learn from cold-start items in the recommendation process. The key is using an item-ID free approach for recommendations. By using a metadata-based representation of items, the M2TRec model learns the representation for items with zero or few interactions. Through experiments on two datasets, we show that M2TRec outperforms several state-of-the-art session-based recommendation models. Multi-task learning contributes to the model's predictive performance. Importantly, M2TRec's core ideas help in generating fast and accurate recommendations for cold start-items, sessions with tail items, and for the task of category prediction. 

\balance
\bibliographystyle{ACM-Reference-Format}
\bibliography{m2trec}

\end{document}